\documentclass[aps,prc,twocolumn]{revtex4}
\usepackage{graphicx}
\usepackage{amssymb}
\usepackage{amsmath}
\begin{document}
\title{Sub-barrier capture with quantum diffusion approach}

\author{V.V.Sargsyan$^1$, R.A.Kuzyakin$^1$, G.G.Adamian$^1$, N.V.Antonenko$^1$, W.Scheid$^2$ and H.Q.Zhang$^3$}
\affiliation{$^{1}$Joint Institute for Nuclear Research, 141980 Dubna, Russia\\
$^{2}$Institut f\"ur Theoretische Physik der Justus--Liebig--Universit\"at, D--35392 Giessen, Germany\\
$^{3}$China Institute of Atomic Energy, Post Office Box 275, Beijing 102413,  China
}
\date{\today}
\begin{abstract}
With the quantum diffusion approach the  behavior of capture cross sections and
mean-square angular momenta of captured systems are revealed in the reactions
with deformed and spherical nuclei
at sub-barrier energies.
With decreasing bombarding energy under the barrier the external turning point
of the nucleus-nucleus potential leaves the region of short-range nuclear
interaction and action of friction.
Because of this  change of the regime of interaction,
an unexpected enhancement of the capture cross
section is found  at bombarding energies far below
the Coulomb barrier.
This effect is shown its worth in the dependence of mean-square angular momentum
 on the bombarding energy.
From the comparison of calculated capture  cross sections and experimental
capture or fusion cross sections
the importance of quasifission near the entrance channel
is demonstrated for the actinide-based reactions  and reactions
with medium-heavy nuclei at extreme sub-barrier energies.
\end{abstract}
\maketitle
\section{Introduction}
\label{intro}
The measurement of excitation functions down to the extreme sub-barrier energy
region is important for  studying the long range behavior
of nucleus-nucleus interaction as well as the coupling of relative motion with other  degrees of freedom
~\cite{BeckermanNi58Ni64,ZhangOth,ScarlassaraNi58Zr94,ZhangOU,Og,ZuhuaFTh,Nadkarni,Mor,trotta,Ji1,Tr1,DasguptaS32Pb208,NishioOU,Ji22,Ji2,NishioSiU,Vino,Dg,Ca2,NishioSU,StefaniniS36Ca48,HindeSTh,Shri,ItkisSU,akn,Nishionew,Siga,MontagnoliS36Ni64}.
The experimental data obtained are of interest
for solving  astrophysical problems related to nuclear synthesis.
Indications for an enhancement of the $S$-factor,
$S=E_{\rm c.m.}\sigma \exp(2 \pi\eta)$~\cite{Zvezda,Zvezda2},
where $\eta(E_{\rm c.m.})=Z_1Z_2e^2\sqrt{\mu/(2\hbar^2E_{\rm c.m.})}$ is the Sommerfeld parameter,
at energies $E_{\rm c.m.}$
below the Coulomb barrier
have been found in Refs.~\cite{Ji1,Ji2,Dg}.
Its origin is still under discussion.

From the comparison of capture cross sections and fusion cross sections
one can show a significant role of the quasifission channel
in the reactions with various medium-light and heavy nuclei at
sub-barrier energies.
The competition  between the complete fusion and quasifission can strongly reduce
the value of the fusion cross section and, respectively,
the value of the evaporation residue cross section ~\cite{Volkov,nasha,Avaz}.
This effect is especially crucial
in the production of superheavy nuclei.
It worth remembering that first evidences of hindrance  for compound nucleus
formation in the reactions with massive nuclei at low energies near the Coulomb barrier
were observed at GSI already long time ago~\cite{GSI}.

To clarify the behavior of capture and fusion cross sections at sub-barrier
energies, a further development of the theoretical methods is required~\cite{Avaz,Gomes,our,Den2}.
The conventional coupled-channel
approach with realistic set of parameters is not able to describe the fusion
cross sections either below or above the Coulomb barrier~\cite{Dg}. The use of a quite
shallow nucleus-nucleus potential~\cite{Es} with an adjusted repulsive core considerably
improves the agreement between the calculated and experimental data.
Besides the coupling with collective excitations,
the dissipation, which is simulated by an imaginary potential in Ref.~\cite{Es} or
by damping in each channel in Ref.~\cite{Hag1}, seems to be important.
The quantum diffusion approach
 based on the quantum master-equation for
the reduced density matrix has been suggested  in Ref.~\cite{EPJSub}
for the describing  capture process.
The collisions of  nuclei are treated in terms
of a single collective variable: the relative distance  between
the colliding nuclei.
Our approach takes into consideration the fluctuation and dissipation effects in
collisions of heavy ions which model the coupling with various channels.
In the present paper the capture model~\cite{EPJSub} is applied.
\begin{figure}
\vspace{-0.cm}
\resizebox{1.0\columnwidth}{!}{
  \includegraphics{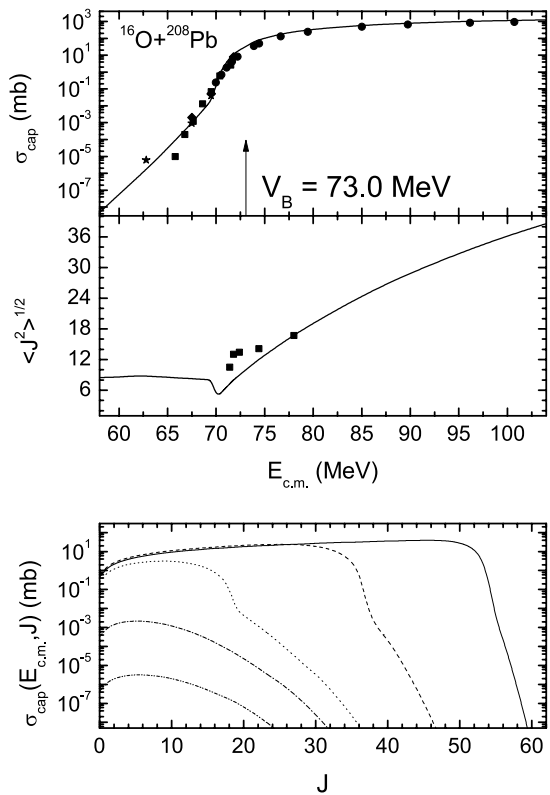} }
\vspace{-0.7cm}
\caption{
The calculated (lines) capture cross section (upper part),
average angular momenta of captured system (middle part) versus $E_{\rm c.m.}$,
and partial capture cross sections (lower part) versus  $J$ at
$E_{\rm c.m.}$=65 (dash-dot-dotted line), 69.5 (dash-dotted line), 73 (dotted line),
83 (dashed line),  and 100 (solid line) MeV
for the $^{16}$O+$^{208}$Pb reaction with the spherical nuclei. The experimental
cross sections marked by closed squares, circles, rhombus,  stars are
from Refs.~\protect\cite{Dg,Mor,Tr1,Og}, respectively. The experimental
values of $\langle J^2\rangle$ (solid squares) are taken from Ref.~\protect\cite{Vand}.
The value of the Coulomb barrier $V_b$ is indicated by arrow.
}
\label{fig:1}       
\end{figure}
\begin{figure}
\resizebox{0.90\columnwidth}{!}{
  \includegraphics{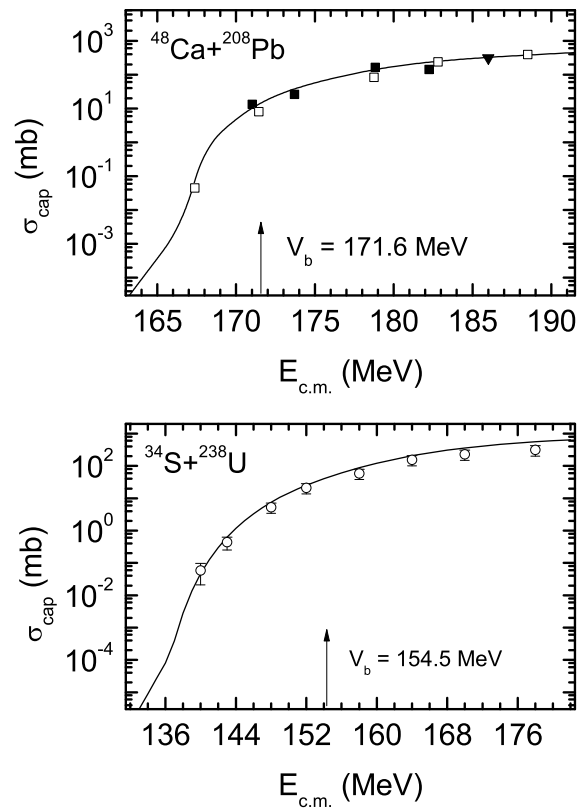} }
\caption{
 The calculated (solid lines) capture cross section versus $E_{\rm c.m.}$
for the reactions $^{48}$Ca+$^{208}$Pb (upper part) and $^{34}$S+$^{238}$U  (lower part).
The experimental cross sections are taken from Refs.~\protect\cite{Ca1}
(closed squares and triangle),~\protect\cite{Ca2} (open squares),
and  ~\protect\cite{Nishionew} (open circles).
The value of the Coulomb barrier $V_b$ is indicated by arrow.
The static quadrupole deformation parameters  are:
$\beta_{1}$($^{48}$Ca)=$\beta_{2}$($^{208}$Pb)=0,
$\beta_{1}$($^{34}$S) =0.125,
and $\beta_{2}$($^{238}$U)=0.286 ~\protect\cite{Ram}.
}
\label{fig:2}       
\end{figure}
\begin{figure}
\vspace{-0.8cm}
\resizebox{0.75\columnwidth}{!}{
  \includegraphics{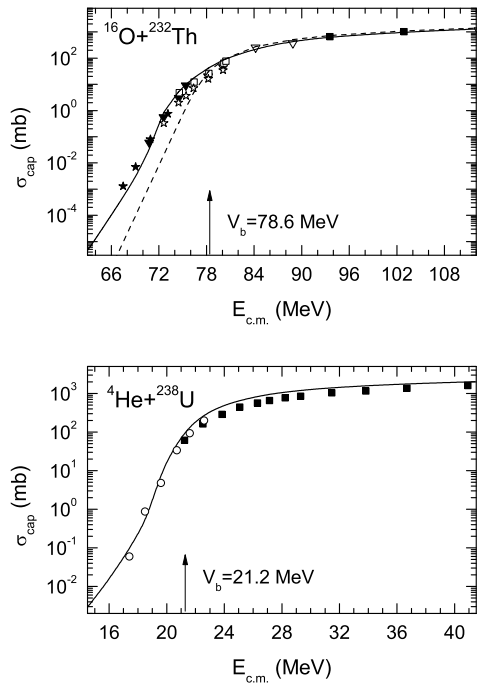} }
\caption{
The same as in Fig.~2, but for the reactions for the reactions
$^{16}$O + $^{232}$Th and $^{4}$He + $^{238}$U.
The experimental  data in the upper part are taken from Refs.~\protect\cite{BackOTh}
(open triangles),~\protect\cite{ZhangOth} (closed triangles),
~\protect\cite{MuakamiOTh} (open squares),~\protect\cite{KailasOTh}
(closed squares),~\protect\cite{ZuhuaFTh} (open stars)
and ~\protect\cite{Nadkarni} (closed stars).
The fission cross sections from Refs.~\protect\cite{trotta}
and ~\protect\cite{ViolaOU} are shown in the lower part
by open circles and solid squares, respectively.
The value of the Coulomb barrier $V_b$ for the spherical nuclei is indicated by arrow.
The dashed curve represents the calculation with the Wong formula ~\protect\cite{Wong}.
The static quadrupole deformation parameters  are: $\beta_{2}$($^{232}$Th)=0.261
$\beta_{2}$($^{238}$U)=0.286 ~\protect\cite{Ram} and
 $\beta_{1}$( $^{4}$He)=$\beta_{1}$($^{16}$O)=0.
}
\label{fig:3}       
\end{figure}
\begin{figure}
\vspace{-0.8cm}
\resizebox{0.75\columnwidth}{!}{
  \includegraphics{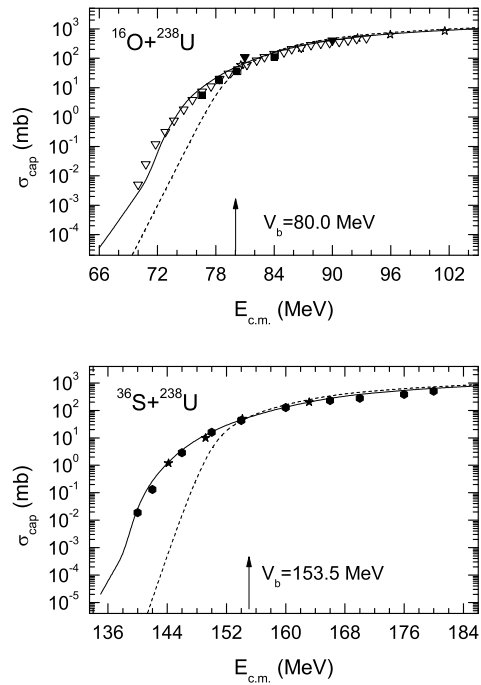} }
\caption{
The same as in Fig.~2, but for the reactions $^{16}$O + $^{238}$U and $^{36}$S + $^{238}$U.
The experimental cross sections are taken from
Refs.~\protect\cite{NishioOU} (open triangles),~\protect\cite{TokeOU} (closed triangles),
~\protect\cite{ZuhuaFTh} (open squares),~\protect\cite{ZhangOU} (closed squares),
~\protect\cite{ViolaOU} (open stars),~\protect\cite{ItkisSU} (closed stars),
and  ~\protect\cite{NishioSU} (rhombuses).
The dashed curve represents the calculation with the Wong formula ~\protect\cite{Wong}.
The static quadrupole deformation parameters  are:
$\beta_{2}$($^{238}$U)=0.286 and
 $\beta_{1}$($^{16}$O)=$\beta_{1}$($^{36}$S)=0.
}
\label{fig:4}       
\end{figure}
\begin{figure}
\vspace{-0.3cm}
\resizebox{0.75\columnwidth}{!}{
  \includegraphics{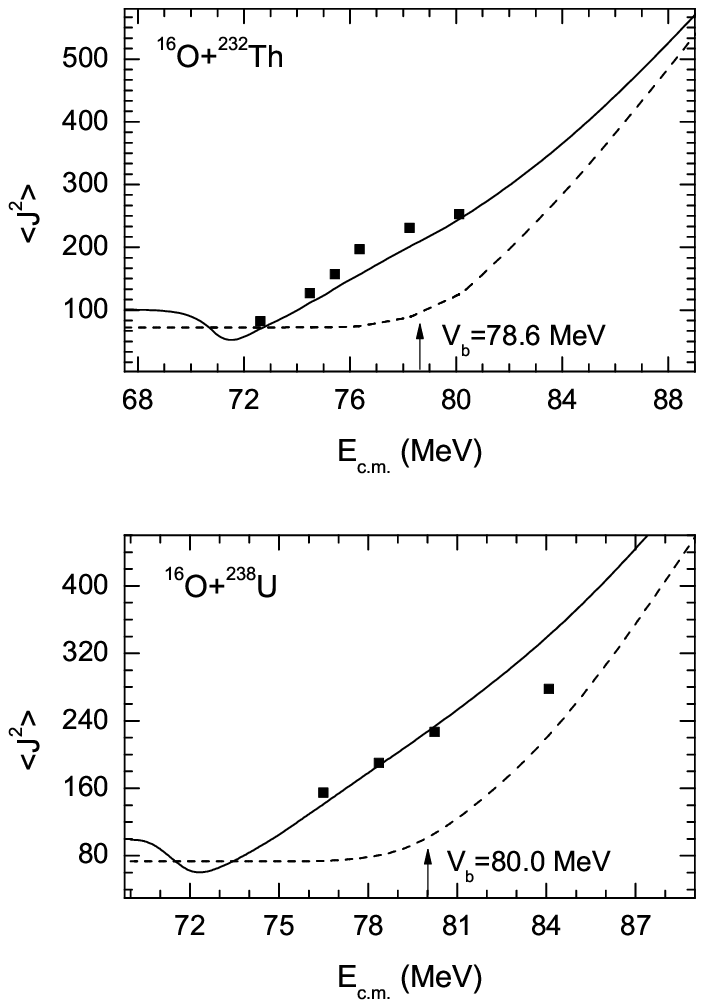} }
\caption{
The calculated mean-square angular momenta  versus $E_{\rm c.m.}$
for the reactions $^{16}$O + $^{232}$Th,$^{238}$U are compared with
experimental data ~\protect\cite{ZuhuaFTh}.
The dashed curve represents the calculation by the Wong-type formula ~\protect\cite{Wong}.
}
\label{fig:5}       
\end{figure}

\section{Comparison with experimental data and predictions}
\label{sec:1}
One can see in
Figs.~1--4 that
with decreasing $E_{\rm c.m.}$ up to about 3 - 18 MeV below the Coulomb barrier
the regime of interaction is changed because at the external
turning point the colliding nuclei do not reach the region of nuclear interaction
where the friction plays a role.
As  result, at smaller $E_{\rm c.m.}$ the capture cross sections $\sigma_{cap}$
fall with a smaller rate.
 Therefore, an
effect of the change of fall rate of sub-barrier capture cross section should be in the data
if we assume that the friction starts to act only when the colliding nuclei approach the barrier.
Note that at sub-barrier energies  the experimental data
have still large uncertainties to make a firm experimental conclusion about
this effect.
The effect seems to be more pronounced in  collisions of spherical nuclei, where
the regime of interaction is changed at $E_{\rm c.m.}$ up to about 3.5 - 5 MeV
below the Coulomb barrier~\cite{EPJSub}.


The calculated  mean-square angular momenta $\langle J^2\rangle$
of captured systems versus $E_{\rm c.m.}$
are presented in Figs.~1 and 5 for the reactions $^{16}$O+$^{208}$Pb
and $^{16}$O + $^{232}$Th,$^{238}$U.
At energies below the barrier  $\langle J^2\rangle$ has a minimum.
This minimum depends on the deformations of nuclei and on the factor $Z_1\times Z_2$. For the
reactions $^{16}$O + $^{232}$Th,$^{238}$U,
these minima are about 7 -- 8 MeV below the corresponding Coulomb barriers, respectively.
The experimental data \cite{Vand} indicate the presence of the minimum as well.
On the left-hand side
of this minimum the dependence of $\langle J^2\rangle$ on $E_{\rm c.m.}$ is rather weak.
Note that the found behavior of $\langle J^2\rangle$, which is
related to the change of the regime of interaction between the colliding nuclei,
would affect the angular
anisotropy of the products of fission-like fragments following capture.
Indeed, the values of $\langle J^2\rangle$
are extracted from  data on angular distribution of fission-like fragments~\cite{akn}.

The agreement between the experimental
$\langle J^2\rangle$ and those calculated with Wong-type formula is rather bad.
At energies below the barrier  the $\langle J^2\rangle$
has no a minimum (see Fig.~5).
However, for the considered reactions the saturation values of $\langle J^2\rangle$
are close to those obtained in our formalism.

In Fig.~6  the calculated astrophysical $S$--factor versus $E_{\rm c.m.}$ is shown
for the  $^{16}$O+$^{238}$U reaction.
The $S$-factor has a maximum which is seen in  experiments \cite{Ji1,Ji2,Es}.
After this maximum $S$-factor slightly decreases with decreasing $E_{\rm c.m.}$
and then starts to increase.
This effect  seems to be more pronounced in  collisions of spherical nuclei.
The same behavior has been revealed in Refs.~\cite{LANG} by extracting the
$S$-factor from the experimental data.

In Fig.~6, the so-called logarithmic
derivative,
$L(E_{\rm c.m.})\break =d(\ln (E_{\rm c.m.}\sigma_{cap}))/dE_{\rm c.m.},$  and
the  barrier distribution\break
 $d^2(E_{\rm c.m.}\sigma_{cap})/d E_{\rm c.m.}^2$ are
presented for the $^{16}$O+$^{238}$U reaction. The logarithmic derivative strongly
increases  below the barrier and then has a maximum at
$E_{\rm c.m.}\approx V_b^{orient}$(sphe-\break re-pole)-3 MeV
(at $E_{\rm c.m.}\approx V_b$-3 MeV for the case of spherical nuclei).
The maximum of $L$ corresponds to the minimum of the $S$-factor.
The  barrier distributions calculated
with an energy increment 0.2 MeV  have only one maximum
at $E_{\rm c.m.}\approx V_b^{orient}$(sphere-sphere)$=V_b$ as in the experiment~\cite{DH}.
With an increasing increment the barrier distribution is  shifted to  lower energies.
Assuming a spherical target nucleus in the calculations,
we obtain a more narrow barrier distribution.
\begin{figure}
\resizebox{0.85\columnwidth}{!}{
  \includegraphics{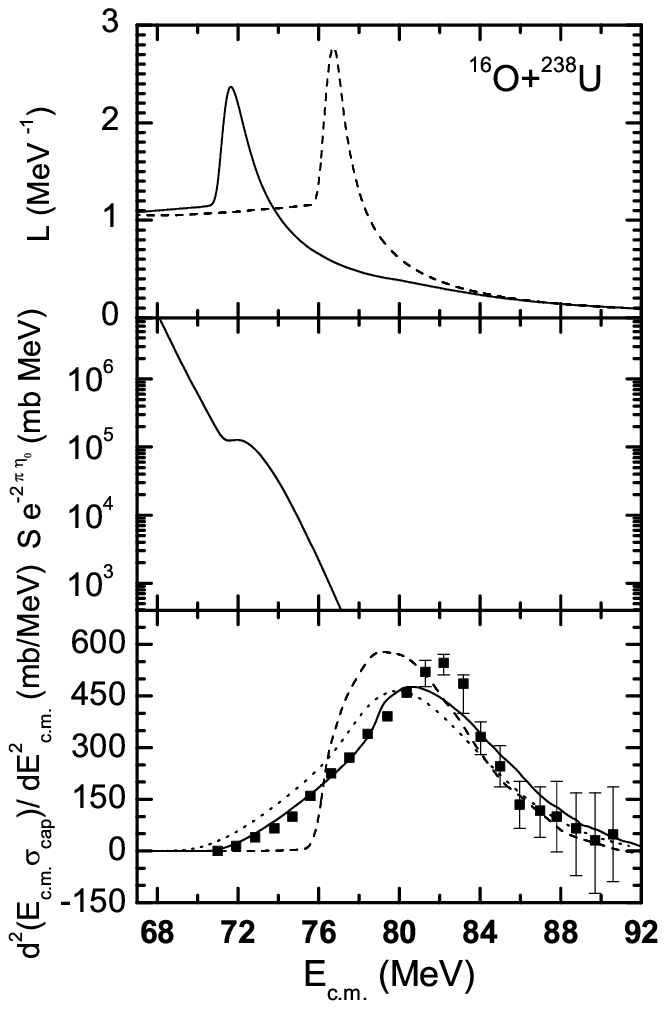} }
\caption{
The calculated values of the
astrophysical $S$-factor with $\eta_0=\eta(E_{\rm c.m.}=V_b)$
(middle part), the logarithmic derivative $L$ (upper part)
and the fusion barrier distribution
$d^2(E_{\rm c.m.}\sigma_{cap})/d E_{\rm c.m.}^2$ (lower part)
for the $^{16}$O+$^{238}$U reaction.
The value of $L$ calculated with the assumption of $\beta_1$($^{16}$O)=$\beta_2$($^{238}$U)=0 is shown by a
dashed line.
The solid and dotted  lines show the values of
$d^2(E_{\rm c.m.}\sigma_{cap})/d E_{\rm c.m.}^2$  calculated
with the increments 0.2 and 1.2 MeV, respectively.
The closed squares are the experimental data of Ref.~\protect\cite{DH}.
}
\label{fig:6}       
\end{figure}
\begin{figure}
\resizebox{0.85\columnwidth}{!}{
  \includegraphics{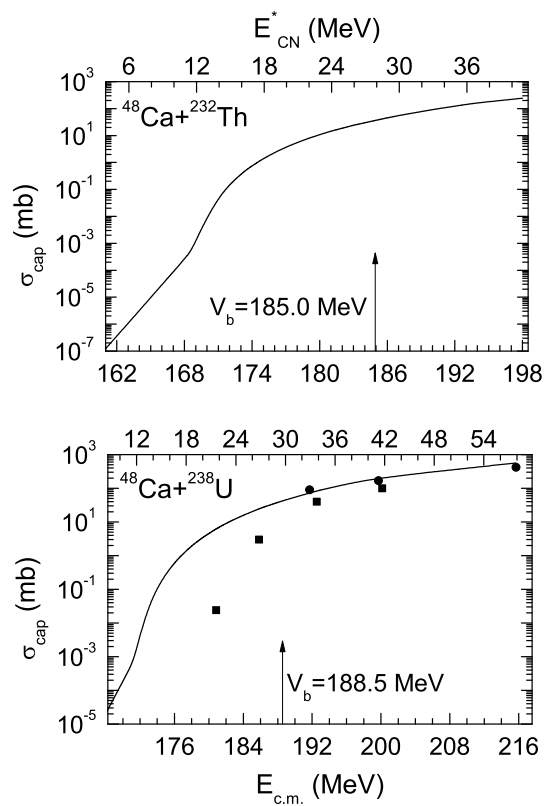} }
\caption{
The same as in Fig.~4, but for the $^{48}$Ca + $^{232}$Th,$^{238}$U reactions.
The excitation energies $E^*_{CN}$ of the corresponding nuclei are indicated.
The experimental data are taken from Refs.~\protect\cite{Itkis1} (marked by squares)
and ~\protect\cite{Shen} (marked by circles).
The static quadrupole deformation parameters  are: $\beta_{2}$($^{238}$U)=0.286,
$\beta_{2}$($^{232}$Th)=0.261,  and
$\beta_{1}$($^{48}$Ca)=0.
}
\label{fig:7}       
\end{figure}

\section{Capture cross sections in reactions with large fraction of quasifission}
\label{sec:2}
In the case of large values of $Z_1\times Z_2$
the  quasifission process  competes with complete
fusion at  energies near barrier and can lead to a large
hindrance for fusion, thus ruling the probability for
producing superheavy elements in the actinide-based reactions~\cite{nasha,trota}.
Since the sum of the fusion cross section $\sigma_{fus}$,
and the quasifission cross section $\sigma_{qf}$ gives the capture cross section,
$$\sigma_{cap}=\sigma_{fus}+\sigma_{qf},$$
and  $\sigma_{fus}\ll \sigma_{qf}$ for
actinide-based reactions
$^{48}$Ca + $^{232}$Th,
$^{238}$U,$^{244}$Pu,$^{246,248}$Cm and $^{50}$Ti + $^{244}$Pu
\cite{nasha}, we have $$\sigma_{cap}\approx\sigma_{qf}.$$

In a wide mass-range near
the entrance channel, the quasifission events overlap with the
products of deep-in-\break elastic collisions and can not be firmly distinguished.
Because of this the mass region near the entrance channel
is taken out in
the experimental analyses in Refs. ~\cite{Itkis1,Itkis2}.
Thus, by comparing the calculated and experimental capture cross sections
one can study the  importance of quasifission near the entrance channel
for the actinide-based reactions leading to  superheavy nuclei.

The capture cross sections for the quasifission reactions \cite{Itkis1,Itkis2,Shen}
are shown in Figs. 7 and 8.
One can observe a large deviations of the experimental data of Refs.~\cite{Itkis1,Itkis2}
from the
the calculated results.
The possible reason is  an
underestimation of  the quasifission yields measured
in these reactions. Thus,
the  quasifission yields near the entrance channel
are important.
Note that there are
the experimental uncertainties in the bombarding energies.

One can see in Fig.~9 that
the experimental  and the theoretical cross sections become closer
with increasing  bombarding energy.
This means that with increasing  bombarding energy
the quasifission yields near the entrance channel mass-region decrease
with respect to the quasifission yields in  other mass-regions.
As seen in Fig.~9,
the quasifission yields near the entrance channel mass-region increase
with $Z_1\times Z_2$.
\begin{figure}
\resizebox{0.85\columnwidth}{!}{
  \includegraphics{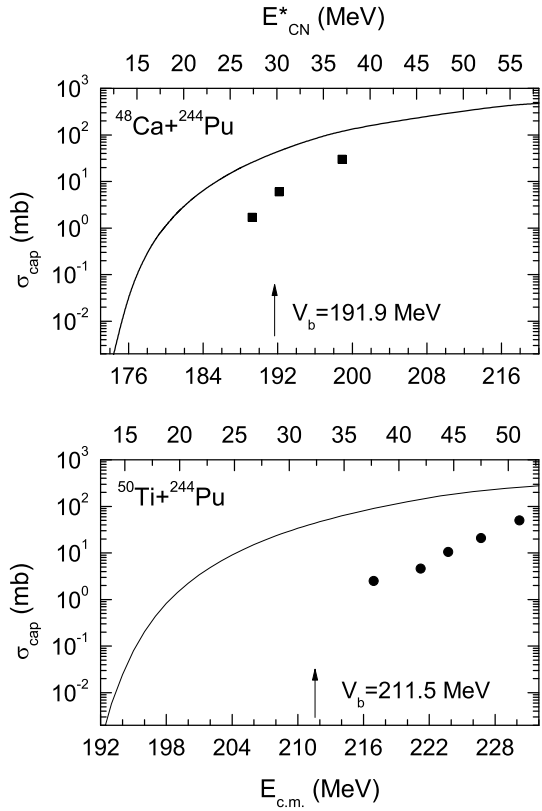} }
\caption{
The same as in Fig.~7, but for the indicated $^{48}$Ca,$^{50}$Ti + $^{244}$Pu reactions.
The experimental data are from Refs.~\protect\cite{Itkis2} (squares)
and ~\protect\cite{Itkis1} (circles).
The static quadrupole deformation parameters  are: $\beta_{2}$($^{244}$Pu)=0.293,
and
$\beta_{1}$($^{48}$Ca)=$\beta_{1}$($^{50}$Ti)=0.
}
\label{fig:8}       
\end{figure}
\begin{figure}
\resizebox{0.85\columnwidth}{!}{
  \includegraphics{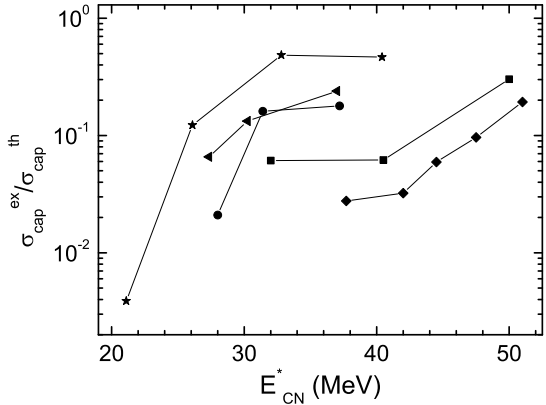} }
\caption{
The ratio of theoretical and experimental capture cross sections versus
the excitation energy $E_{\rm c.m.}$ of the compound nucleus
for the  reactions $^{48}$Ca+$^{238}$U (closed stars),
$^{48}$Ca+$^{244}$Pu (closed triangles),
$^{48}$Ca+$^{246}$Cm (closed squares),
$^{48}$Ca+$^{248}$Cm (closed circles),  and $^{50}$Ti+$^{244}$Pu (closed rhombuses).
}
\label{fig:9}       
\end{figure}
\begin{figure}
\vspace{0.6cm}
\resizebox{0.85\columnwidth}{!}{
  \includegraphics{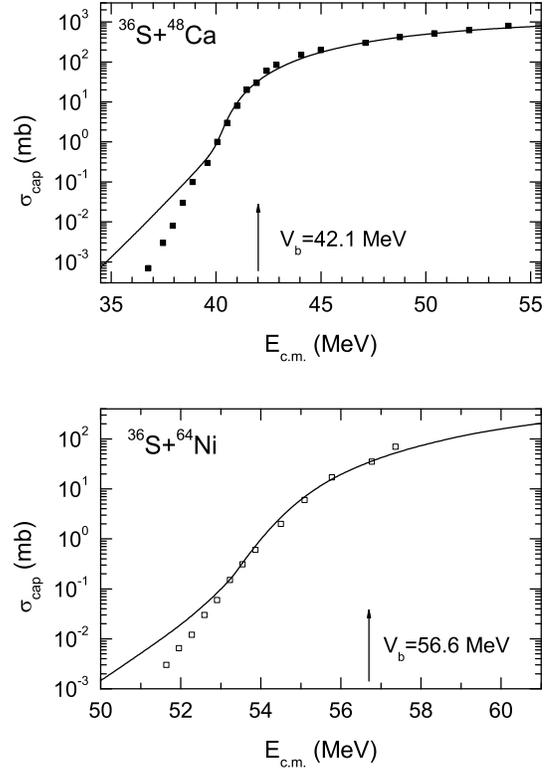} }
\caption{
The calculated capture cross sections versus $E_{\rm c.m.}$ for the indicated reactions.
The experimental fusion cross sections marked by closed and open squares are taken from
Refs.~\protect\cite{StefaniniS36Ca48,MontagnoliS36Ni64}, respectively.
The values of the Coulomb barrier are indicated by arrows.
The static quadrupole deformation parameters  are:
$\beta_{1}$($^{36}$S)=$\beta_{2}$($^{48}$Ca)=0 and $\beta_{2}$($^{64}$Ni)=0.087~\protect\cite{Moel1}.
}
\label{fig:10}       
\end{figure}
\begin{figure}
\vspace{-0.cm}
\resizebox{0.9\columnwidth}{!}{
  \includegraphics{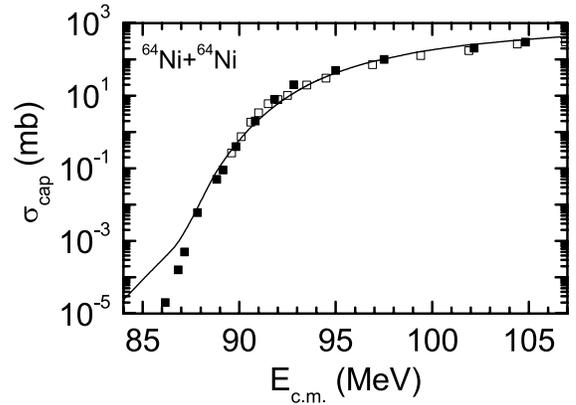} }
\vspace{-0.cm}
\caption{
The calculated capture cross sections versus $E_{\rm c.m.}$ for the indicated reaction.
The experimental fusion cross sections marked by closed  squares and  circles
are taken from Refs.~\protect\cite{BeckermanNi58Ni64,Ji22}, respectively.
Here, $\beta_{1,2}$($^{64}$Ni)=0.087.
}
\label{fig:11}       
\end{figure}
\section{Origin of fusion hindrance in  reactions with medium-mass nuclei at  sub-barrier energies}
\label{sec:3}
In Figs.~10 and 11 the calculated capture cross section are presented for
the reactions $^{36}$S + $^{48}$Ca,$^{64}$Ni and $^{64}$Ni + $^{64}$Ni.
The values of $V_b$ are adjusted to the experimental data for the fusion cross sections  shown as well.
For the systems mentioned above,
the difference between the sub-barrier capture and fusion cross sections
becomes larger with decreasing bombarding energy $E_{\rm c.m.}$.
The same effect one can see for the $^{16}$O + $^{208}$Pb  reaction.
Assuming that the estimated capture  and the measured fusion cross sections are correct,
the small fusion cross section at energies well below the Coulomb barrier
may indicate that other reaction channel is open and
the system evolves by other reaction mechanism after the capture.
The observed hindrance factor may be understood in term of quasifission
whose cross section should be added to the one of
fusion to obtain a meaningful comparison with the calculated capture cross section.
The quasifission event  corresponds to the formation
of a nuclear-molecular state or dinuclear system
with small excitation energy that separates
(in the competition with  the  compound nucleus formation process)
by the quantal tunneling
through the Coulomb barrier
 in a binary event with mass and charge  close
to the entrance channel. In this sense the quasifission is the general phenomenon
which takes place in the reactions with the
massive~\cite{Volkov,nasha,Avaz,GSI}, medium-mass  and,  probably, light nuclei.
%
%
%
For the medium-mass  and light nuclei,
this  reaction mode has to be studied in the future experiments:
from the mass (charge) distribution measurements one can
show the distinct components due to quasifission.
The low-energy experimental data would probably provide straight information
since the high-energy data may be shaded by competing reaction
processes such as quasifission and
deep-inelastic collisions.
Note that the binary decay events were already observed experimentally in \cite{Henning}
for the $^{58}$Ni + $^{124}$Sn reaction at energies below the Coulomb barrier but
assumed to be related to deep-inelastic scattering.

\section{Summary}
\label{sec:4}
The quantum diffusion approach is applied to study
the capture process in the reactions with  deformed and spherical nuclei at sub-barrier energies.
Due to a change of the regime of interaction (the turning-off of the nuclear forces and friction)
at sub-barrier energies, the curve related to the  capture cross section
as a function of bombarding energy has smaller slope.
This change is also reflected in the functions
$\langle J^2\rangle$, $L(E_{\rm c.m.})$, and $S(E_{\rm c.m.})$.
The mean-square angular momentum of captured system versus $E_{\rm c.m.}$ has a minimum
and then saturates at sub-barrier energies.
This behavior of $\langle J^2\rangle$ would increase the expected anisotropy
of the angular distribution of the products of fission and quasifission following capture.
The astrophysical factor has a maximum and a minimum at energies below the barrier.
The maximum of $L$-factor corresponds to the minimum of the $S$-factor.
The importance of quasifission near the entrance channel
is shown for the actinide-based reactions at near barrier energies
and reactions with medium-mass nuclei
at extreme sub-barrier energies.
One can suggest the experiments to check these predictions.

We thank H.~Jia and D.~Lacroix for fruitful discussions and  suggestions.
We are grateful to K.~Nishio for providing us his experimental data.
This work was supported by DFG, NSFC, and RFBR.
The IN2P3-JINR and Polish-JINR Cooperation
programs are gratefully acknowledged.\\

\end{document}